\documentclass[twocolumn,aps,pra,tightenlines,showpacs,superscriptaddress,10pt]{revtex4-1}

\usepackage[colorlinks=true,urlcolor=blue,citecolor=blue,linkcolor=blue,bookmarks=false]{hyperref} %
\usepackage{amsmath,amssymb,mathrsfs,amsfonts,bm}
\usepackage{graphicx}
\usepackage{dcolumn}
\usepackage{tabularx}
\usepackage{color}
\usepackage{cleveref}

\begin{document}


\title{Magnetic-impurity resonance states for different pairing symmetries in twisted bilayer graphene}

\author{Liang Chen}
\email[Corresponding Email: ]{slchern@ncepu.edu.cn}
\affiliation{School of Mathematics and Physics, North China Electric Power University, Beijing, 102206, China}

\author{Hui-Zhen Li}
\affiliation{School of Mathematics and Physics, North China Electric Power University, Beijing, 102206, China}

\author{Rong-Sheng Han}
\affiliation{School of Mathematics and Physics, North China Electric Power University, Beijing, 102206, China}

\date{\today}

\begin{abstract}
  In this work, we study the magnetic-impurity resonance states in superconducting phase of twisted bilayer graphene for different pairing symmetries. Using the two-orbital model proposed by Yuan and Fu in
  {[\emph{Phys. Rev. B} \textbf{98}, 045103 (2018)]}, we find that when the impurity is located at one site of the emergent honeycomb lattice, the spacial distributions of the resonance states will break both the threefold and twofold rotation symmetries of $D_3$ group for pairing symmetries which belong to the irreducible representations of this point group. When the magnetic impurity is located at the center of the emergent honeycomb lattice, the appearance of resonance peak at the position close to the impurity can be considered as a strong evidence of non-$s$-wave pairing.
\end{abstract}


\maketitle


\section{Introduction} \label{sec1} 

Recent experiments \cite{Cao2018Nature1,Cao2018Nature2} on bilayer graphene with a small twisted angle have attracted great attention. Theoretical investigations \cite{Santos2007PRL,Bistritzer2011PNAS,Morell2010PRB,Laissardiere2012PRB,FangS2016PRB} show that when the twisted angle is close to the `magic' angles, the bilayer graphene exhibits some low energy flat bands in the mini Brillouin zone (MBZ) of the moir{\'e} patten superlattice. Experimental measurement \cite{Cao2018Nature2} demonstrates that these flat bands display the insulating phase at half-filling, i.e., the charge density is tuned to $\pm2e$ per unit cell of the superlattice via gate voltage. When the charge density is doped slightly away from these correlated insulating phases, superconducting phases are observed.  For a record-low carrier density of a few $10^{11}$ cm$^{-2}$, a remarkable high critical temperature $T_c=1.7$K is observed \cite{Cao2018Nature1}.
The strong correlation features in this material are very similar to the high-$T_c$ cuprate superconductors. Which makes graphene moir\'e superlattice to be an experimental tunable platform for unconventional superconductivity with completely carbon atoms.

The nature of the superconductivity phases and the mechanism for the relative high critical temperature in `magic angle' twisted bilayer graphene (TwBG) are not clear so far. Theoretical investigations from strong correlation physics show that the unconvertional pairings are preferred in this material, i.e., the topological nontrivial $d$+$id$ pairing is repeatedly proposed in theoretical works\cite{XuC2018arXiv1803,GuoH2018arXiv1803,HuangT2018arXiv1804,LiuCC2018arXiv1804,KennesDM2018arXiv1805,Fidrysiak2018arXiv1805}, the $p$-wave pairing \cite{RoyB2018arXiv1803}, $s^{\pm}$ and $s^{++}$ pairing \cite{Sherkunov2018arXiv1807} are also candidates dependent on different conditions. Some other theoretical investigations show that the TwBG exhibits strong electron-phonon coupling at certain twisted angles and electron densities, which leads to a remarkable high critical temperature of magnitude $1\sim10$K \cite{WuFC2018arXiv1805,LianB2018arXiv1807,Peltonen2018arXiv1805}. As demonstrated in these works, the pairing potential  mediated by electron-phonon coupling  reveals the $s$-wave symmetry. Further investigations show that, if the electron-electron interaction is sufficient large, the $d$-wave symmetric pairing potential is also possible for the electron-phonon coupling mechanism  \cite{WuFC2018arXiv1805}. Among the comprehensive investigations on the nature of superconductivity in this material, the determination of pairing symmetry is of fundamental importance.

In this work, we study the magnetic impurity resonance states for typical pairing symmetries in the superconductivity phases of TwBG. Like the standard phase-sensitive tetracrystal measurements \cite{Tsuei1994PRL,Tsuei1997Nature} and quasiparticle interference (QPI) experiments \cite{Hoffman2002Science,WangQH2003PRB,Hanaguri2007NatPhys,Hanaguri2009Science}, the local density of states (LDOS) of the resonance states near a magnetic (and nonmagnetic) impurity is an important method to uncover the symmetries of pairing potentials in unconventional superconductors. This method is widely used in the high-$T_c$  cuprate superconductors \cite{PanSH2000Nat,HudsonEW2001Nat,BobroffJ2001PRL,ZhangGM2001PRL,Polkovnikov2001PRL,ZhuJX2000PRB,VojtaM2002PRB,Polkovnikov2002PRB,DaiX2003PRB,BaarS2016JSNM}, iron-based superconductors \cite{TsaiWF2009PRB,BangY2009PRB,AkbariA2010PRB}, chiral $p$-wave superconductors and topological superconductors \cite{SauJD2013PRB,FuZG2012JPCM,ZhaGQ2017EPL,GuoYW2017FP}, etc.

The paper is organized as follows: In Sec. {\ref{sec2}}, we propose the model Hamiltonian with four different pairing symmetries and give a theoretical derivation of the LDOS. In Sec. {\ref{sec3}}, we analysis the numerical results for the pairing symmetries. According to different representations of the $D_3$ group, the LDOS for three kinds of different hybridizations are presented. A conclusion is given in Sec. {\ref{sec4}}.

\section{Theoretical Model}\label{sec2} 

\begin{figure}[tb]
	\includegraphics[width=\columnwidth]{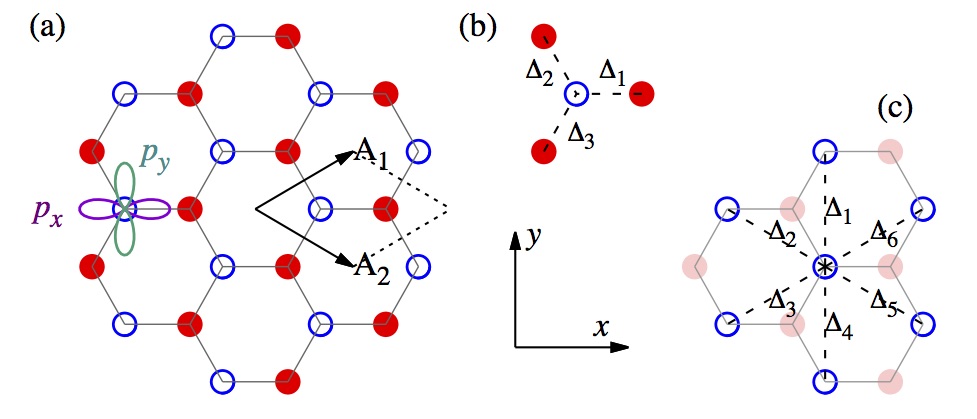}
	\caption{(color online) (a) Schematic representation of the two-orbital tight-binding model. The two primitive vectors are chosen as $\bm{A}_{1,2}=A(\sqrt{3}/2,\pm1/2)$, where $A$ is the superlattice constant, which is set to be unit length in this work. The red dots and blue circles represent the two sublattices corresponding to the AB and BA spots of the moir{\'e} patten. On each site of the emergent honeycomb lattice, there is a set of $p_x$- and $p_y$-like Wannier orbitals which belongs to the $\mathcal{E}$ representation of the $D_3$ point group (see Tab \ref{tab1} for the irreducible representations of $D_3$ group). The original point is set to be located at the impurity site. (b) The nearest neighbor pairing. For the extended s-wave pairing, the potential is set to be $\Delta_1=\Delta_2=\Delta_3=\Delta_{\rm{s^{\prime}}}$. For the $p$-wave pairing, the potential is set to be $\Delta_1=\Delta_p$, $\Delta_2=e^{i\phi}\Delta_p$, $\Delta_3=e^{2i\phi}\Delta_p$ with $\phi=2\pi/3$. (c) The next nearest neighbor pairing. For the \emph{d}+\emph{id}-wave pairing, the potential is set as $\Delta_1=\Delta_4=\Delta_{d+id}$, $\Delta_2=\Delta_5=e^{i\phi}\Delta_{d+id}$, $\Delta_3=\Delta_6=e^{2i\phi}\Delta_{d+id}$.}%
	\label{fig1}%
\end{figure}

\begin{figure}[t]
	\includegraphics[width=\columnwidth]{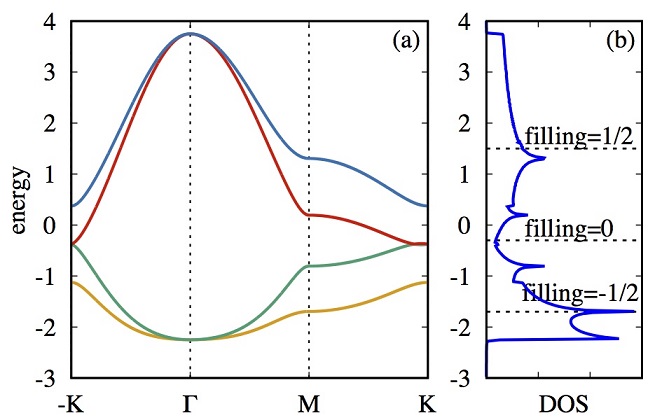}
	\caption{(color online) (a) Band structure of the tight-binding model Hamiltonian (\ref{eq1}) for $t_1=1$, $t_2=t_1/9$, $t_1^{\prime}=t_1/4$, $t_2^{\prime}=t_2/4$, $t_5^{\prime\prime}=0$ and $\mu=0$. (b) The corresponding DOS. The dashed lines show the filling ratio, the specified filling ratios $1/2$ and $-1/2$ are close to the Van Hove singularities. }%
	\label{fig2}%
\end{figure}

\begin{figure}[t]
	\includegraphics[width=\columnwidth]{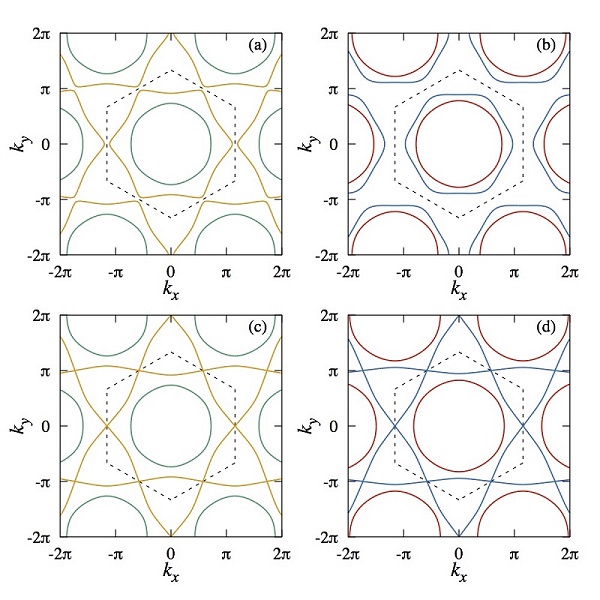}
	\caption{(color online) Fermi surface for different chemical potentials. (a) $\mu=-1.7$ (filling ratio $=-1/2$). (b) $\mu=1.5$ (filling ratio $=1/2$). (c) $\mu=-1.6945$ (the Van Hove singularity). (d) $\mu=1.3058$ (the other Van Hove singularity). The dashed line labels the MBZ. The colors for different pockets of the Fermi surface is specified to corresponding to the colors of band structure shown in Fig. \ref{fig2}(a).}%
	\label{fig3}%
\end{figure}

We use the two-orbital tight-binding model proposed by Yuan and Fu \cite{YuanNFQ2018PRB} with a few modifications to describe the low-energy physics of TwBG, which is expressed in the following form,
\begin{gather}
H_0=\sum_{\bm{k}}\psi_{\bm{k}}^{\dag}h(\bm{k})\psi_{\bm{k}}, \label{eq1} \\
h(\bm{k})=\sum_{i,j}\epsilon_{i,j}(\bm{k})\tau_i\otimes\chi_j,  \label{eq2}
\end{gather}
where $i,j=0,x,y,z$, $\tau_0$ and $\chi_0$ are the $2\times2$ identity matrices in the emergent AB-BA sublattice  space and $\{{p_x,p_y}\}$ orbital space, respectively. $\tau_{x,y,z}$ and $\chi_{x,y,z}$ are the corresponding Pauli matrices in these spaces. $\psi_{\bm{k}}=(c_{{\rm AB},p_x,\bm{k}},c_{{\rm AB},p_y,\bm{k}},c_{{\rm BA},p_x,\bm{k}},c_{{\rm BA},p_y,\bm{k}})^{\mathsf{T}}$, $c_{\alpha,o,\bm{k}}$ is the annihilation operator of electron state for specified sublattice $\alpha={\rm AB}, {\rm BA}$, orbital $o=p_x, p_y$, and wave-vector $\bm{k}$ (the spin index is suppressed here).  The superscript $\mathsf{T}$ means matrix transpose. As the Hermitian conjugate of $\psi_{\bm{k}}$, $\psi_{\bm{k}}^{\dag}$ is the creation operator of corresponding electron state. The summation in Eq. (\ref{eq1}) is taken over the MBZ of TwBG. Fig. \ref{fig1}(a) shows the primitive vectors, unit cell, and $p_{x,y}$-orbitals of the emergent superlattice. The explicit expressions of $\epsilon_{i,j}(\bm{k})$ are given in the following forms,
\begin{gather}
\epsilon_{x,0}(\bm{k})=t_1\left(\cos\frac{k_x}{\sqrt{3}}+2\cos\frac{k_x}{2\sqrt{3}}\cos\frac{k_y}{2}\right), \label{eq3} \\
\epsilon_{y,0}(\bm{k})=t_1\left(\sin\frac{k_x}{\sqrt{3}}-2\sin\frac{k_x}{2\sqrt{3}}\cos\frac{k_y}{2}\right), \label{eq4} \\
\epsilon_{0,0}(\bm{k})=(2t_2+t_2^{\prime})\left(2\cos\frac{\sqrt{3}k_x}{2}\cos\frac{k_y}{2}+\cos{k_y}\right)-\mu, \label{eq5} \\
\epsilon_{x,z}(\bm{k})=t_1^{\prime}\left(\cos\frac{k_x}{\sqrt{3}}-\cos\frac{k_x}{2\sqrt{3}}\cos\frac{k_y}{2}\right), \label{eq6} \\
\epsilon_{y,z}(\bm{k})=t_1^{\prime}\left(\sin\frac{k_x}{\sqrt{3}}+\sin\frac{k_x}{2\sqrt{3}}\cos\frac{k_y}{2}\right),  \label{eq7} \\
\epsilon_{x,x}(\bm{k})=-\sqrt{3}t_1^{\prime}\sin\frac{k_x}{2\sqrt{3}}\sin\frac{k_y}{2},  \label{eq8} \\
\epsilon_{y,x}(\bm{k})=-\sqrt{3}t_1^{\prime}\cos\frac{k_x}{2\sqrt{3}}\sin\frac{k_y}{2},  \label{eq9} \\
\epsilon_{0,z}(\bm{k})=t_2^{\prime}\left(\cos\frac{\sqrt{3}k_x}{2}\cos\frac{k_y}{2}-\cos{k_y}\right),  \label{eq10} \\
\epsilon_{0,x}(\bm{k})=-\sqrt{3}t_2^{\prime}\sin\frac{\sqrt{3}k_x}{2}\sin\frac{k_y}{2},  \label{eq11}  \\
\epsilon_{0,y}(\bm{k})=-t_5^{\prime\prime}\sin\frac{\sqrt{3}k_x}{2}\sin\frac{\sqrt{3}k_x-3k_y}{4}\sin\frac{\sqrt{3}k_x+3k_y}{4},  \label{eq12}
\end{gather}
where $\mu$ is the Fermi energy, $t_1$ is the real hopping amplitude between nearest neighbors within different sublattices, $t_2$ denotes the real hopping amplitude between next nearest neighbors within the same sublattice. $t_5^{\prime\prime}$ represents the inter-sublattice hopping amplitude between fifth nearest neighbors with definite chirality. This term breaks the emergent $SU(4)$ symmetry and hence the fourfold degeneracy (orbital and spin) along the $\Gamma$M line in the MBZ is split. The terms proportional to $t_1^{\prime}$ and $t_2^{\prime}$ refer to the nearest neighbor and next to nearest neighbor hopping amplitudes when the inter-orbital scattering is included. These terms further break the orbital $U(1)$ symmetry and the fourfold degeneracy along the K$\Gamma$ and MK lines in band structure. It is easy to check that this Hamiltonian preserves all the symmetries proposed in Ref. \cite{YuanNFQ2018PRB}. Tab. \ref{tab1} shows the group elements and irreducible representations of the point group $D_3$. For the current Hamiltonian, the specified representations of the group elements are present in the caption of Tab. \ref{tab1}. Fig. \ref{fig2} shows the band structure and density of states (DOS) of the model Hamiltonian with proper parameters. The three dashed lines in Fig. \ref{fig2}(b) from top to bottom denote the typical charge density $+2e$, $0$ and $-2e$ in unit cell of the superlattice, respectively. One can find that filling ratios $1/2$ and $-1/2$ are close to the two Van Hove singularities of the band structure. Fig. \ref{fig3} shows the Fermi surfaces of these two filling ratios ($\pm1/2$) and the near-by Van Hove singularities for the given parameters. When the relative strong electron-electron interactions are presented, TwBG can be driven into the Mott insulate phase at these two half-filling ratios. The experiment \cite{Cao2018Nature1} shows that the superconducting phase appears when the electron concentration is doped slightly away from these half-filling ratios.

\begin{table}[tb]
  \centering
  \begin{tabular}{|c|c|c|c|c|c|c|}
    \hline
    $D_3$ & $\mathscr{E}$ & $2\mathscr{C}_3$ & $3\mathscr{C}^{\prime}$ & linear & quadratic & cubic \\
    \hline
    $\mathcal{A}_1$ & $+1$ & $+1$ & $+1$ & - & $x^2+y^2$ & $x(x^2-3y^2)$\\
    \hline
    $\mathcal{A}_2$ & $+1$ & $+1$ & $-1$ & - & - & $y(3x^2-y^2)$\\
    \hline
    $\mathcal{E}$ & $+2$ & $-1$ & $0$ & $(x,y)$ & $(x^2-y^2,xy)$ & $(x^3+xy^2,x^2y+y^3)$ \\
    \hline
  \end{tabular}
  \caption{Character tab for point group $D_3$. $\mathcal{A}_1$, $\mathcal{A}_2$ and $\mathcal{E}$ in the first column represent the three irreducible representations. $\mathscr{E}$ in the first row refers to the identity element of the $D_3$ point group. $2\mathscr{C}_3$ means the two group elements related to threefold rotation operations along the $z$-axis, under the basis shown in the context below Eq. (\ref{eq2}), these group elements are given by, $\mathscr{C}_z=\tau_0\otimes(-\frac{1}{2}\chi_0-i\frac{\sqrt{3}}{2}\chi_y)$ and the square of $\mathscr{C}_z$, which is also the inverse of $\mathscr{C}_z$, denoted as $\mathscr{C}_z^2$. $3\mathscr{C}^{\prime}$ in the first row means the three group elements related to the twofold rotation operations along the three symmetric axes, under the basis shown in the context below Eq. (\ref{eq2}), these group elements are given by $\mathscr{C}^{\prime}_y=-\tau_x\otimes\chi_z$, $\mathscr{C}_z\mathscr{C}^{\prime}_y\mathscr{C}_z^{-1}$, and $\mathscr{C}_z^2\mathscr{C}^{\prime}_y\mathscr{C}_z^{-2}$. The last three columns show the exemplified polynomial representations.}
  \label{tab1}
\end{table}

Generally, the pairing potential term can be written as,
\begin{equation}
H_{\rm{pairing}}=\sum_{\bm{k},s,s^{\prime}}\psi_{-\bm{k},s}^{\mathsf{T}}\Delta^{s,s^{\prime}}(\bm{k})\psi_{\bm{k},s^{\prime}}+h.c., \label{eq13}
\end{equation}
where $s,s^{\prime}=\uparrow, \downarrow$ refer to the spins of electron states, $h.c.$ means Hermitian conjugate, $\Delta^{s,s^{\prime}}(\bm{k})$ is the pairing potential matrix.
In this work, we consider only the spin-singlet pairing (potential spin-triplet $p$-wave and $d$+$id$ pairing are discussed in Refs. \cite{Isobe2018arxiv1805,Fidrysiak2018arXiv1805}), such that the summation over spin indices contains only one situation, $s=\uparrow$ and $s^{\prime}=\downarrow$. Without causing confusion, we suppress the superscript of pairing potential. In this case, $\Delta(\bm{k})$ is a $4\times4$ matrix in the orbital and sublattice spaces. Following the previous studies, we consider four different pairing potentials in this work. The first one is the on-site $s$-wave pairing,
\begin{equation}
\Delta_s(\bm{k})=\Delta_s\tau_0\otimes\chi_0, \label{eq14}
\end{equation}
the second one is the nearest neighbor (extended) $s$-wave pairing,
\begin{gather}
\Delta_{s'}(\bm{k})=\Delta_{s'}\left[\left(\cos\frac{k_x}{\sqrt{3}}+2\cos\frac{k_x}{2\sqrt{3}}\cos\frac{k_y}{2}\right)\tau_x\otimes\chi_0\right. \notag\\
+\left.\left(\sin\frac{k_x}{\sqrt{3}}-2\cos\frac{k_y}{2}\sin\frac{k_x}{2\sqrt{3}}\right)\tau_y\otimes\chi_0\right],  \label{eq15}
\end{gather}
the third one is the nearest neighbor $p$-wave pairing,
\begin{gather}
\Delta_{p}(\bm{k})=\Delta_p\left[\left(\cos\frac{k_x}{\sqrt{3}}-\cos\frac{k_x}{2\sqrt{3}}\cos\frac{k_y}{2}\right.\right. \notag \\
\left.+i\sqrt{3}\sin\frac{k_x}{2\sqrt{3}}\sin\frac{k_y}{2}\right)\tau_x\otimes\chi_0+\left(\sin\frac{k_x}{\sqrt{3}}\right. \notag \\
\left.\left.+\sin\frac{k_x}{2\sqrt{3}}\cos\frac{k_y}{2}+i\sqrt{3}\cos\frac{k_x}{2\sqrt{3}}\sin\frac{k_y}{2}\right)\tau_y\otimes\chi_0\right],  \label{eq16}
\end{gather}
the last one is the $d$+$id$ pairing between next to nearest neighbor sites,
\begin{gather}
\Delta_{d+id}(\bm{k})=2\Delta_{d+id}\left(\cos{k_y}-\cos\frac{k_y}{2}\cos\frac{\sqrt{3}k_x}{2}\right. \notag \\
\left.+i\sqrt{3}\sin\frac{k_y}{2}\sin\frac{\sqrt{3}k_x}{2}\right)\tau_0\otimes\chi_0,  \label{eq17}
\end{gather}
where the constants $\Delta_s$, $\Delta_{s^{\prime}}$, $\Delta_{p}$ and $\Delta_{d+id}$ represent the strength of the pairing potentials for these four different symmetries. It is easy to check that the $s$-wave and extended $s$-wave pairing potentials belong to the one-dimensional $\mathcal{A}_1$ representation of the $D_3$ group, the $p$-wave and $d$+$id$ pairing potentials belong to the two-dimensional representation (the $\mathcal{E}$ representation shown in Tab. \ref{tab1}) of the $D_3$ group. Figs. \ref{fig1}(b) and \ref{fig1}(c) show the nearest neighbor and the next to nearest neighbor pairing potentials in superlattice space.

\begin{figure}[tb]
	\includegraphics[width=\columnwidth]{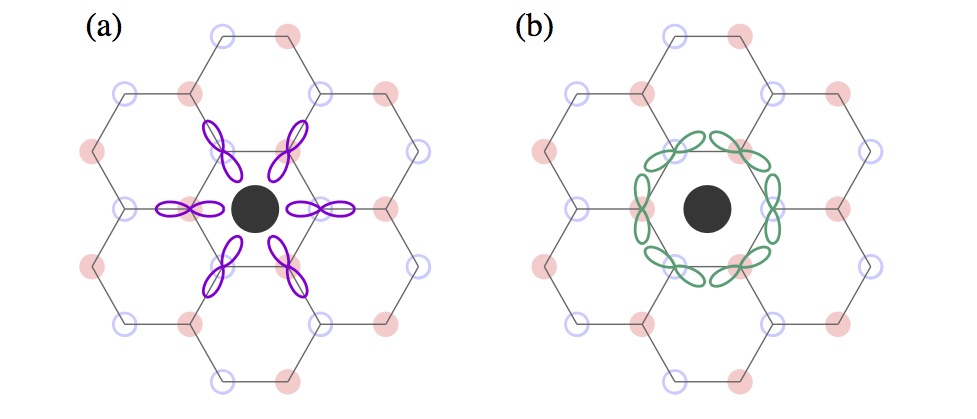}
	\caption{(color online) The two different hybridizations belong to the two-dimensional representation of the $D_3$ group. The magnetic impurity (labeled with black dot) is located at the center of the emergent honeycomb lattice. The orientations of the orbital labels refer to proper combinations of the $p_{x,y}$-orbitals at specified sites. (a) The real space representation of the 1st hybridization $\bar{V}_{\bm{k}}^{(1)}$ given in Eq. (\ref{eq21}). (b) The real space representation of the 2nd hybridization $\bar{V}_{\bm{k}}^{(2)}$ given in Eq. (\ref{eq22}).}%
	\label{fig4}%
\end{figure}

The total Hamiltonian for a magnetic impurity coupled to the superconducting TwBG is consisted of four terms,
\begin{gather}
H=H_{\rm{imp}}+H_{\rm{hyb}}+H_0+H_{\rm{pairing}},  \label{eq18} \\
H_{\rm{imp}}=\sum_{s}\epsilon_dd_s^{\dagger}d_s+Ud_{\uparrow}^{\dag}d_{\uparrow}d_{\downarrow}^{\dag}d_{\downarrow},  \label{eq19} \\
H_{\rm{hyb}}=\frac{1}{\sqrt{N}}\sum_{\bm{k},s}\left(\psi_{\bm{k},s}^{\dag}\bar{V}_{\bm{k}}d_{s}+h.c.\right),  \label{eq20}
\end{gather}
where $\epsilon_d$ is the impurity energy, $U$ is the on-site Coulomb interaction, $d_{s}^{\dag}$ and $d_s$ are the creation and annihilation operators of spin-$s$ impurity state, respectively. $N$ is the total number of wave-vectors in the MBZ. $\bar{V}_{\bm{k}}$ is the $4\times1$ hybridization matrix between the impurity state and the conduction state with wave-vector $\bm{k}$. This term has to be considered very carefully. Generally, the impurity Hamiltonian (\ref{eq19}) belongs to the $\mathcal{A}_1$ representation of the $D_3$ group, and the two Wannier orbitals belong to the two-dimensional representation of the $D_3$ group. If the impurity is coupled to only one site of the emergent honeycomb lattice,
i.e., shown by Eq. (\ref{eq23}) in the following context,
the hybridization term (\ref{eq20}) will break both the threefold and twofold rotation symmetries. Firstly, we consider the $\mathscr{C}_3$ rotation symmetric hybridizations, we consider that the impurity is located at the center of the emergent honeycomb lattice, i.e., the center of an AA spot of the moir\'e pattern. The impurity is hybridized to the six nearest neighbors symmetrically, detailed analysis show that there are only two different hybridizations belong to the two-dimensional representation of $D_3$ group, Fig. \ref{fig4} shows the form of these hybridizations in real space.
Their explicit expressions in wave-vector space are given by,
\begin{gather}
\bar{V}_{\bm{k}}^{(1)}=V_0\left(
\begin{matrix}
2\left(e^{i\frac{k_x}{\sqrt{3}}}-e^{-i\frac{k_x}{2\sqrt{3}}}\cos\frac{k_y}{2}\right) \\
2\sqrt{3}i e^{-i\frac{k_x}{2\sqrt{3}}}\sin\frac{k_y}{2}\\
-2\left(e^{-i\frac{k_x}{\sqrt{3}}}-e^{i\frac{k_x}{2\sqrt{3}}}\cos\frac{k_y}{2}\right) \\
2\sqrt{3}i e^{i\frac{k_x}{2\sqrt{3}}}\sin\frac{k_y}{2}
\end{matrix}
\right),  \label{eq21} \\
\bar{V}_{\bm{k}}^{(2)}=V_0\left(
\begin{matrix}
2\sqrt{3}i e^{-i\frac{k_x}{2\sqrt{3}}}\sin\frac{k_y}{2}\\
-2\left(e^{i\frac{k_x}{\sqrt{3}}}-e^{-i\frac{k_x}{2\sqrt{3}}}\cos\frac{k_y}{2}\right) \\
-2\sqrt{3}i e^{i\frac{k_x}{2\sqrt{3}}}\sin\frac{k_y}{2} \\
-2\left(e^{-i\frac{k_x}{\sqrt{3}}}-e^{i\frac{k_x}{2\sqrt{3}}}\cos\frac{k_y}{2}\right) \\
\end{matrix}
\right),  \label{eq22}
\end{gather}
where $V_0$ represents the strength of the hybridizations.
When the impurity is located at the center of an AB spot and hybridized to the $p_x$ and $p_y$ orbitals at this emergent site equally, the hybridization matrix in Eq. (\ref{eq20}) can be written as,
\begin{equation}
\bar{V}_{\bm{k}}^{(3)}=(V_0, V_0, 0, 0)^{\mathsf{T}}. \label{eq23}
\end{equation}
We need to emphasize that, in this work, the original point is set to be located at the impurity site, which means that, for the hybridizations $\bar{V}_{\bm{k}}^{(1)}$ and $\bar{V}_{\bm{k}}^{(2)}$, the original point is located at the center of the emergent honeycomb lattice, for the hybridization $\bar{V}_{\bm{k}}^{(3)}$, the original point is located at the center of one of the AB spot of the moir\'e pattern.

Now we consider the solution of the total Hamiltonian (\ref{eq18}), in the strong Coulomb interaction limit, $U\rightarrow\infty$, the double occupation state on the impurity site can be excluded, this condition may be represented by introducing the slave-boson  operators $b$ and $b^{\dag}$, $d_s=f_sb^{\dag}$, $d_s^{\dag}=bf_s^{\dag}$. The extra degrees of freedom can be ruled out by the no-double occupation condition, $Q=b^{\dag}b+\sum_{s}f_s^{\dag}f_s=I$. In the mean-field approximation, the slave-boson operators can be replaced by the expectation value $\langle{b}\rangle=\langle{b^{\dag}}\rangle=b_0$, the constraint condition can be approximated by introducing a Lagrangian multiplier term to the Hamiltonian, $\lambda_0(b_0^2+\sum_sf_s^{\dag}f_s-1)$. The parameters $b_0$ and $\lambda_0$ can be determined by minimizing the free energy of the mean-field Hamiltonian. This mean-field Hamiltonian is given by,
\begin{gather}
H_{\rm{MF}}=H_{\rm{imp}}^{\rm{MF}}+H_{\rm{hyb}}^{\rm{MF}}+H_0+H_{\rm{pairing}}+\lambda_0(b_0^2-1),  \label{eq24} \\
H_{\rm{imp}}^{\rm{MF}}=\sum_{s}\tilde{\epsilon}_df_s^{\dagger}f_s,    \label{eq25} \\
H_{\rm{hyb}}^{\rm{MF}}=\frac{1}{\sqrt{N}}\sum_{\bm{k},s}\psi_{\bm{k},s}^{\dag}\tilde{\bar{V}}_{\bm{k}}d_{s}+h.c.,  \label{eq26}
\end{gather}
where $\tilde{\epsilon}_d=\epsilon_d+\lambda_0$ is the renormalized impurity energy, $\tilde{\bar{V}}_{\bm{k}}=b_0\bar{V}_{\bm{k}}$ is the renormalized hybridization matrix. In the Bogoliubov-de Gennes (BdG) formalism, the mean-field Hamiltonian can be recast as,
\begin{gather}
H_{\rm{BdG}}=\lambda_0(b_0^2-1)+\tilde{\epsilon}_d+\Phi^{\dag}\Lambda\Phi+\sum_{\bm{k}}\Psi_{\bm{k}}^{\dag}h_{\rm{BdG}}(\bm{k})\Psi_{\bm{k}}  \notag \\
+\frac{1}{\sqrt{N}}\sum_{\bm{k}}\left[\Psi_{\bm{k}}^{\dag}\mathcal{V}_{\bm{k}}\Phi+h.c.\right] \label{eq27}
\end{gather}
where the Nambu spinor is defined as, $\Psi_{\bm{k}}^{\dag}=(\psi_{\bm{k},\uparrow}^{\dag},\psi_{-\bm{k},\downarrow}^{\mathsf{T}})$, $\Phi=(f_{\uparrow},f_{\downarrow}^{\dag})^{\mathsf{T}}$, $\Lambda=\tilde{\epsilon}_d\tau_0\otimes\chi_0\otimes\varsigma_z$,
\begin{gather}
h_{\rm{BdG}}(\bm{k})=\left( \begin{matrix}
h(\bm{k}) &  \Delta(\bm{k}) \\
\Delta^{\dag}(\bm{k}) & -h(\bm{k})
\end{matrix}\right), \mathcal{V}_{\bm{k}}=\left(\begin{matrix}
\tilde{\bar{V}}_{\bm{k}} & 0 \\
0 & -\tilde{\bar{V}}_{\bm{k}} \end{matrix}\right), \label{eq28}
\end{gather}
$\varsigma_z$ is the third Pauli matrix in the Nambu spinor space.

Using the standard functional integration method in quantum field theory, we find that the finite temperature free energy of the magnetic impurity can be written as,
\begin{gather}
\mathcal{F}=\lambda_0(b_0^2-1)+\tilde{\epsilon}_d+k_BT\sum_{n}{\rm{Tr}}\ln[G_f(i\omega_n)],  \label{eq29}
\end{gather}
where $k_B$ is the Boltzmann constant and $T$ refers to temperature, $\omega_n=(2n+1){\pi}k_BT$ is the Matsubara frequency of fermion, $G_f(i\omega_n)=[i\omega_n-\Lambda-\Sigma_f(i\omega)]^{-1}$ is the Green's function of impurity state expressed in imaginary frequency representation, $\Sigma_f(i\omega_n)$ is the self-energy of the impurity state. It is given by, $\Sigma_f(i\omega_n)=\frac{1}{N}\sum_{\bm{k}}\mathcal{V}^{\dag}_{\bm{k}}G_{\psi}^{(0)}(i\omega_n,\bm{k})\mathcal{V}_{\bm{k}}$, where $G_{\psi}^{(0)}(i\omega_n,\bm{k})=[i\omega_n-h_{\rm{BdG}}(\bm{k})]^{-1}$ is the unperturbed Green's function of the conduction states in superconducting TwBG.  The parameters $\lambda_0$ and $b_0$ are determined by minimizing the free energy (\ref{eq29}), which gives,
\begin{gather}
b_0^2+k_BT\sum_{n}{\rm Tr}\left[G_{f}(i\omega_n)\varsigma_z\right]=0,  \label{eq30} \\
\lambda_0b_0^2+k_BT\sum_{n}{\rm Tr}\left[G_{f}(i\omega_n)\Sigma_{f}(i\omega_n)\right]=0,  \label{eq31}
\end{gather}

The LDOS near the magnetic impurity is obtained by analytic continuation of the imaginary-time Green's function, $i\omega_n\rightarrow{E+i0^{+}}$,
\begin{gather}
\rho_{\psi}(E,\bm{R})=-\frac{1}{\pi}{\rm Im}{\rm Tr}\left[G_{\psi}(E;\bm{R},\bm{R})\frac{1+\varsigma_z\otimes\tau_0\otimes\chi_0}{2}\right], \label{eq32}
\end{gather}
where $G_{\psi}(E;\bm{R},\bm{R})$ is the full Green's function of the conduction states, which is given by \cite{ZhangGM2001PRL},
\begin{gather}
G_{\psi}(E;\bm{R},\bm{R})=\frac{1}{N}\sum_{\bm{k},\bm{k^{\prime}}}e^{i(\bm{k}-\bm{k^{\prime}})\cdot\bm{R}}G_{\psi}(E;\bm{k},\bm{k^{\prime}}),  \label{eq33} \\
G_{\psi}(E;\bm{k},\bm{k^{\prime}})=G_{\psi}^{(0)}(E,\bm{k})\left[\delta_{\bm{k},\bm{k^{\prime}}}+\mathcal{T}_{\bm{k},\bm{k^{\prime}}}(E)G_{\psi}^{(0)}(E,\bm{k^{\prime}})\right], \label{eq34}
\end{gather}
where $\mathcal{T}_{\bm{k},\bm{k^{\prime}}}(E)=\mathcal{V}_{\bm{k}}G_{f}(E)\mathcal{V}^{\dag}_{\bm{k^{\prime}}}/N$ is the $\mathcal{T}$-matrix.

\begin{figure*}[tb]
	\includegraphics[width=\textwidth]{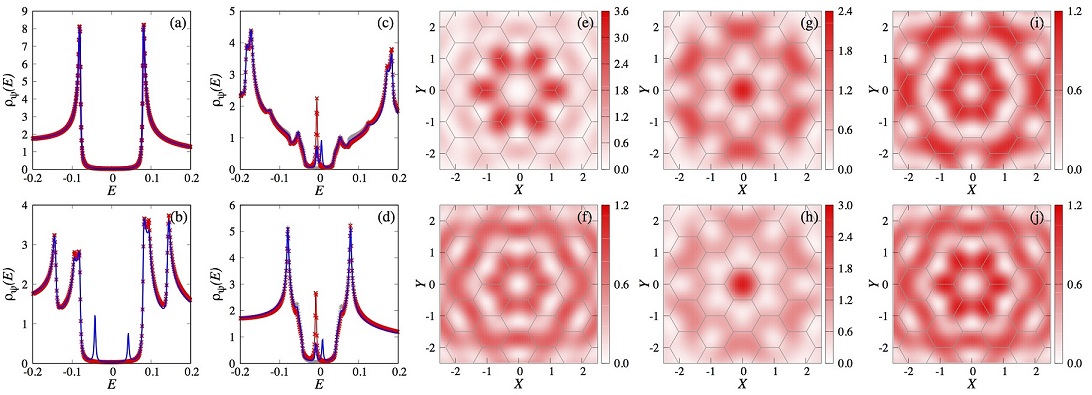}
	\caption{(color online) LDOS for the case with the first hybridization $\bar{V}_{\bm{k}}^{(1)}$, the impurity is located at the center of the emergent honeycomb lattice, $\bm{R}=(0,0)$. (a)-(d) LDOS as a function of energy for $s$-wave, extended $s$-wave, $p$-wave and $d$+$id$ pairing, respectively. The gray lines labeled with `$+$' in (a)-(d) show the clean LDOS without impurity. The red lines labeled with `$\times$' in (a)-(d) show the LDOS near the impurity $\bm{R}=(0,0)$. The blue lines in (a)-(d) show the LDOS at the nearest neighbor site $\bm{R}=(-1/\sqrt{3},0)$. (e) and (f) show the real space distributions of the LDOS corresponding to the left ($E<0$) and right ($E>0$) in-gap resonance peaks of the blue line in (b), respectively (extended $s$-wave pairing). (g) and (i) show the real space distributions of the LDOS corresponding to the two in-gap resonance peaks of the blue line in (c), respectively ($p$-wave pairing). (h) and (j) present the real space distributions of the LDOS of the two in-gap resonance peaks for $d$+$id$ pairing shown in (d).  The silvery lines in (e)-(j) show the emergent honeycomb lattice. The parameters are chosen as follows: $\Delta_s=\Delta_{s^{\prime}}=\Delta_p=0.08$, $\Delta_{d+id}=0.02$, $V_0=\Delta_s$, $\epsilon_d=-\Delta_s/4$.}%
	\label{fig5}%
\end{figure*}

\begin{figure*}[tb]
	\includegraphics[width=0.8\textwidth]{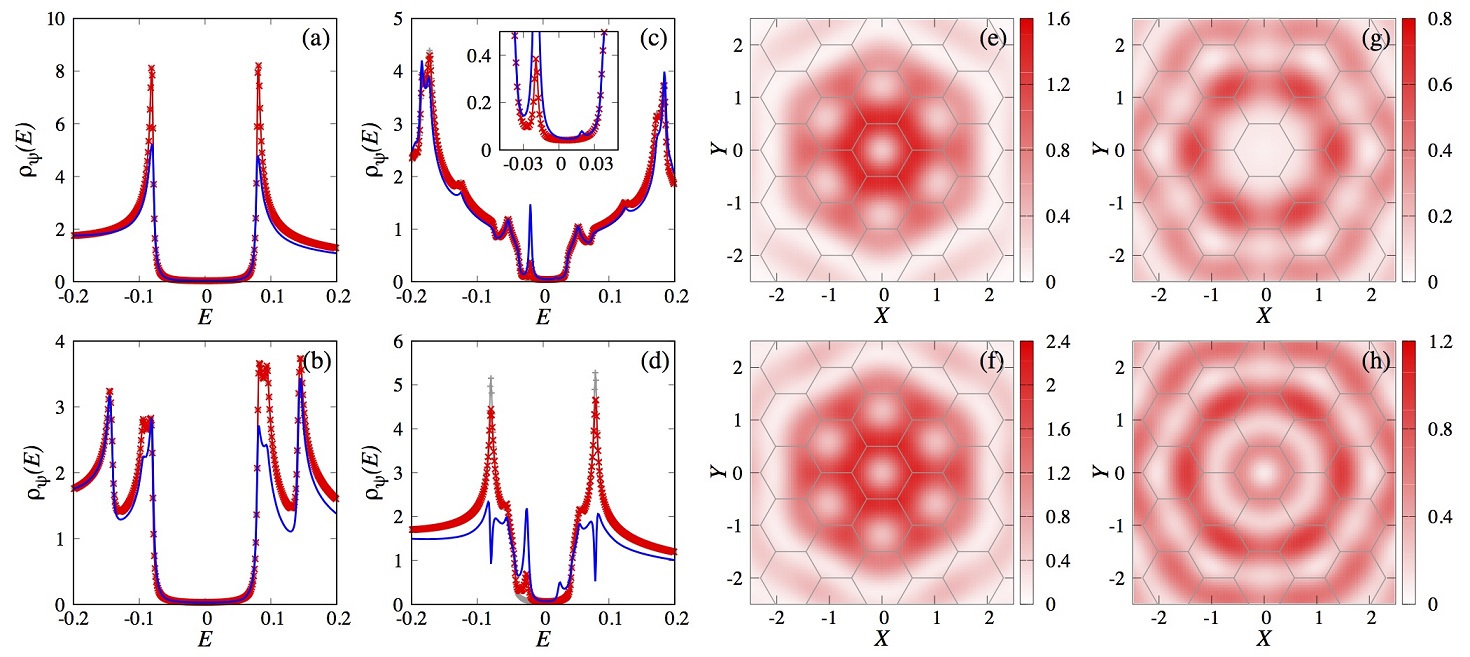}
	\caption{(color online) LDOS for the case with the second hybridization $\bar{V}_{\bm{k}}^{(2)}$, the impurity is located at the center of the emergent honeycomb lattice, $\bm{R}=(0,0)$. (a)-(d) LDOS as a function of energy for $s$-wave, extended $s$-wave, $p$-wave and $d$+$id$ pairing, respectively. The gray lines labeled with `$+$' in (a)-(d) show the clean LDOS without impurity. The red lines labeled with `$\times$' in (a)-(d) show the LDOS near the impurity $\bm{R}=(0,0)$. The blue lines in (a)-(d) show the LDOS at the nearest neighbor site $\bm{R}=(-1/\sqrt{3},0)$. The insert in (c) shows the details of the resonance peaks. (e) and (g) show the real space distributions of the LDOS corresponding to the two in-gap resonance peaks of the blue line in (c), respectively ($p$-wave pairing). (f) and (h) present the real space distributions of the LDOS of the two in-gap resonance peaks for $d$+$id$ pairing shown in (d). The silvery lines in (e)-(j) show the emergent honeycomb lattice. The parameters are chosen as follows: $\Delta_s=\Delta_{s^{\prime}}=\Delta_p=0.08$, $\Delta_{d+id}=0.02$, $\epsilon_d=-\Delta_s/4$, $V_0=\Delta_s$ for the $s$-wave (a) and extended $s$-wave (b) pairing, $V_0=\Delta_s/4$ for the $p$-wave pairing, and $V_0=2\Delta_s$ for the $d$+$id$ pairing (d).}%
	\label{fig6}%
\end{figure*}

\section{Numerical results}\label{sec3}

Before the detailed discussion of the numerical results, we need to clarify that, as demonstrated in previous works \cite{Withoff1990PRL,Cassanello1996PRB,Cassanello1997PRB,GonzalezBuxton1998PRB,ZhangGM2001PRL,ZhuangHB2009EPL,UchoaB2012PRL}, there are two distinct phases for magnetic doping in superconductor and marginal fermi liquid. When the hybridization between the impurity and host material is not strong enough, the impurity behaviors as an isolated impurity and decoupled to the host material, as a consequence, Eqs. (\ref{eq30})-(\ref{eq31}) have no solution. For the other situation, the impurity and the host material are strong correlated, the in-gap Yu-Shiba-Rusinov \cite{Yu1965APS,Shiba1968CSS,Rusinov1969JETPL} resonance states present. Hereafter, the hybridization and impurity energy are chosen in the strong correlation regime such that we can get the in-gap resonance states.
In the practical calculation, a mesh size of $3\times1024\times1024$ in the momentum space is chosen. The infinitesimal imaginary part of the retarded Green's function has been set to be $1.0\times10^{-3}$.

Now we consider the LDOS for charge density equals to $-2e$ in a unit cell of the moir\'e pattern superlattice,
$\mu=-1.7$. Figs. \ref{fig5}(a)-\ref{fig5}(d) show the LDOSs as functions of energy at some specified positions in real space for hybridization $\bar{V}_{\bm{k}}^{(1)}$ and different symmetries. In this case, the impurity (the original point) is located at the center of the emergent honeycomb lattice. The red lines labeled with `$\times$' and solid blue lines present the LDOSs for  $\bm{R}=(0,0)$ and $\bm{R}=(-1/\sqrt{3},0)$, respectively, the gray lines labeled with `{\scriptsize$+$}' refer to the clean LDOSs without impurity. For the $s$-wave pairing (Fig. \ref{fig5}(a)) and extended $s$-wave pairing (Fig. \ref{fig5}(b)), one can find that the in-gap resonance peak at $\bm{R}=(0,0)$ is vanishing, we need to emphasize that this is not because $V_0$ is too small to couple the impurity and the superconductor, actually, for the given parameters $V_0=\Delta_s$ and $\epsilon_d=-\Delta_s/4$, the numerical calculations give non-zero solutions of $b_0$ and $\lambda_0$. Furthermore, detailed numerical analysis show that the red line labeled with `$\times$' and the gray line labeled with `$+$' in Fig. \ref{fig5}(a) (and also Fig. \ref{fig5}(b)) coincide to each other exactly. These observations lead to a central conclusion in this work.
Here we demonstrate from symmetry analysis that \emph{the non-vanishing resonance peak at the center of the AA spot (the center of the emergent honeycomb lattice) can be considered as a strong evidence of the unconventional pairings belong to the two-dimensional representation or the $\mathcal{A}_2$ representation of the $D_3$ group}. Generally, the in-gap resonance peak is induced by the impurity scattering term in Eq. (\ref{eq34}). If the pairing potential belongs to the $\mathcal{A}_1$ representation (the identity representation), the BdG Hamiltonian (\ref{eq27}) will be invariant under the $D_3$ group operations, i.e., $\mathscr{C}_3H_{\rm BdG}\mathscr{C}_3^{-1}=H_{\rm BdG}$ and $\mathscr{C}^{\prime}H_{\rm BdG}\mathscr{C}^{\prime -1}=H_{\rm BdG}$, so that $\mathscr{C}_3G_{\psi}^{(0)}(E,\bm{k})\mathscr{C}_3^{-1}=G_{\psi}^{(0)}(E,\mathscr{C}_3^{-1}\bm{k}\mathscr{C}_3)$, $\mathscr{C}^{\prime}G_{\psi}^{(0)}(E,\bm{k})\mathscr{C}^{\prime -1}=G_{\psi}^{(0)}(E,\mathscr{C}^{\prime -1}\bm{k}\mathscr{C}^{\prime})$. Here $\mathscr{C}_3$ and $\mathscr{C}^{\prime}$ denote the threefold and twofold rotation operations given in Tab \ref{tab1}. The hybridizations $\bar{V}_{\bm{k}}^{(1)}$ and $\bar{V}_{\bm{k}}^{(2)}$ preserve the desired symmetries, $\mathscr{C}_3\bar{V}_{\bm{k}}^{(1,2)}=\bar{V}_{\mathscr{C}_3^{-1}\bm{k}\mathscr{C}_3}^{(1,2)}$, $\mathscr{C}^{\prime}\bar{V}_{\bm{k}}^{(1,2)}=\bar{V}_{\mathscr{C}^{\prime -1}\bm{k}\mathscr{C}^{\prime}}^{(1,2)}$, so we get,
\begin{gather}
\mathscr{C}_3{\sum_{\bm{k}}G_{\psi}^{(0)}(E,\bm{k})\mathcal{V}_{\bm{k}}}\mathscr{C}_3^{-1}=\sum_{\bm{k}}G_{\psi}^{(0)}(E,\bm{k})\mathcal{V}_{\bm{k}},  \label{eq35} \\
\mathscr{C}^{\prime}{\sum_{\bm{k}}G_{\psi}^{(0)}(E,\bm{k})\mathcal{V}_{\bm{k}}}\mathscr{C}^{\prime -1}=\sum_{\bm{k}}G_{\psi}^{(0)}(E,\bm{k})\mathcal{V}_{\bm{k}},  \label{eq36}
\end{gather}
it is easy to check that the only solution of these equations is $\sum_{\bm{k}}G_{\psi}^{(0)}(E,\bm{k})\mathcal{V}_{\bm{k}}=0$. Substituting this result into Eqs. (\ref{eq34}) and (\ref{eq33}), one can find that, at the original point, $\bm{R}=(0,0)$, the scattering term has no contribution.  This analysis is evident for the $s$-wave pairing and extended $s$-wave pairing which belong to the $\mathcal{A}_1$ representation. For the topological nontrivial $p$-wave pairing and $d$+$id$ pairing, one can check that $\mathscr{C}_z\Delta_{p}(\bm{k})\mathscr{C}_z^{-1}=e^{-2\pi{i}/3}\Delta_{p}(\mathscr{C}_z^{-1}\bm{k}\mathscr{C}_z)$, $\mathscr{C}_z\Delta_{d+id}(\bm{k})\mathscr{C}_z^{-1}=e^{2\pi{i}/3}\Delta_{d+id}(\mathscr{C}_z^{-1}\bm{k}\mathscr{C}_z)$, such that the in-gap resonance peaks may be non-vanishing, as evident in Figs. \ref{fig5}(c) and \ref{fig5}(d), there does exhibit one resonance peak below the fermi energy for each case (see the red lines labeled with `$\times$').

Previous theoretical investigations show that, both the quantum fluxes \cite{IvanovDA2001PRL} and magnetic impurities \cite{HuH2013PRL} can induce Majorana bound states in topological superconductors. We need to emphasize that the unique in-gap resonance peak can not be regarded as the Majorana bound state because the resonance peak is not located at $E=0$. In addition, the numerical results show that the other resonance peak with opposite energy appears at $\bm{R}\ne(0,0)$, i.e., the blue lines in Figs. \ref{fig5}(c) and \ref{fig5}(d) give the results for LDOS versus energy at the nearest site of the impurity, $\bm{R}=(-1/\sqrt{3},0)$, one can find that there are two resonance peaks in the superconducting gap. These results demonstrate that there do exist two in-gap resonance energy and they are not protected by particle-hole symmetry.
The other related issue we need to emphasize is that the absence of resonance peak at $\bm{R}=(0,0)$ does not mean that the resonance state is absent. The blue line in Fig. \ref{fig5}(b) shows LDOS versus energy at $\bm{R}=(-1/\sqrt{3},0)$, one can find that there are two in-gap resonance peaks symmetric located at the two sides of the Fermi energy. Figs. \ref{fig5}(e) and \ref{fig5}(f) show the corresponding spacial distributions of these two resonance states. These two patterns are significantly different from each other, which demonstrate that the particle-hole symmetry is broken. Figs. \ref{fig5}(g), \ref{fig5}(i) and \ref{fig5}(h), \ref{fig5}(j) show the spacial distributions of the resonance states near the magnetic impurity for the $p$-wave pairing potential and the $d$+$id$ pairing potential, respectively. One can find that all of these patterns reveal the sixfold rotation symmetry. The maximum intensity is located at $\bm{R}=(0,0)$ in Figs. \ref{fig5}(g) and \ref{fig5}(h), which show the spacial distributions of LDOSs for the negative resonance energies. For the other cases, the maximum intensity is located at the bonds linking the nearest and next to the nearest neighbors of the impurity.

\begin{figure}[t]
	\includegraphics[width=\columnwidth]{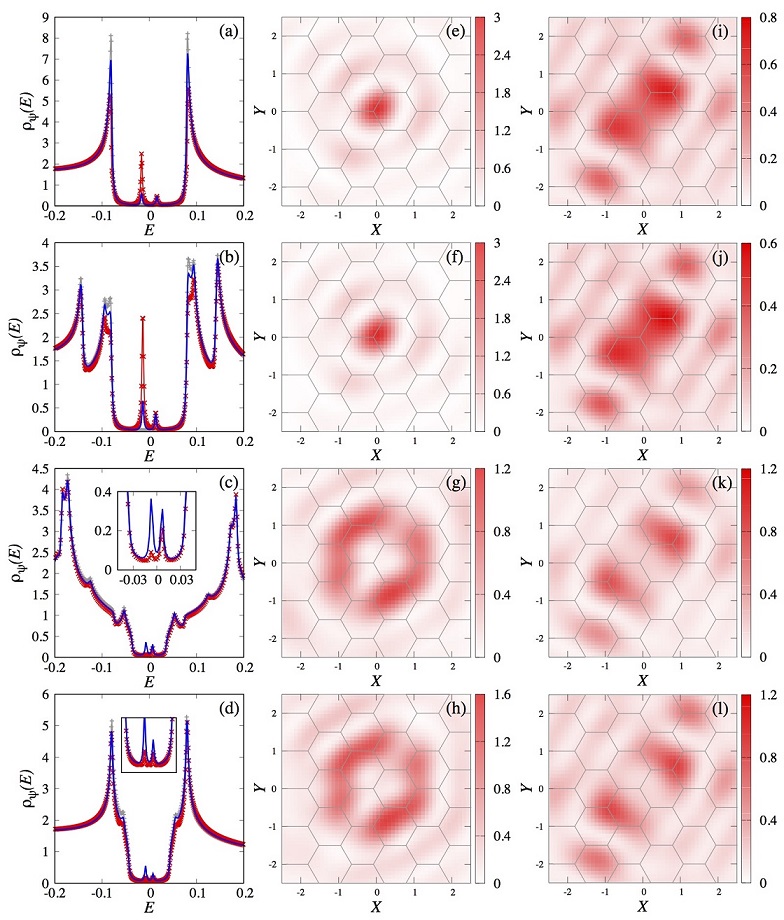}
	\caption{(color online) LDOS for the case with the third hybridization $\bar{V}_{\bm{k}}^{(3)}$, the impurity is located at the center of one of the AB spot, $\bm{R}=(0,0)$. (a)-(d) LDOS as a function of energy for $s$-wave, extended $s$-wave, $p$-wave and $d$+$id$ pairing, respectively. The gray lines labeled with `$+$' in (a)-(d) show the clean LDOS without impurity. The red lines labeled with `$\times$' in (a)-(d) show the LDOS near the impurity $\bm{R}=(0,0)$. The blue lines in (a)-(d) show the LDOS at the nearest neighbor site $\bm{R}=(-1/\sqrt{3},0)$. The second and third columns show the spacial distributions of the LDOS corresponding to the two in-gap resonance peaks shown in the first column for each row.
The inserts in (c) and (d) show the details of the resonance peaks. The silvery lines in (e)-(k) show the emergent honeycomb lattice. The parameters are chosen as follows: $\Delta_s=\Delta_{s^{\prime}}=\Delta_p=0.08$, $\Delta_{d+id}=0.02$, $\epsilon_d=-\Delta_s/4$, $V_0=\Delta_s$ for the $s$-wave (a) and extended $s$-wave (b) pairing, $V_0=\Delta_s/4$ for the $p$-wave pairing, and $V_0=2\Delta_s$ for the $d$+$id$ pairing (d).}%
	\label{fig7}%
\end{figure}

\begin{figure}[t]
	\includegraphics[width=\columnwidth]{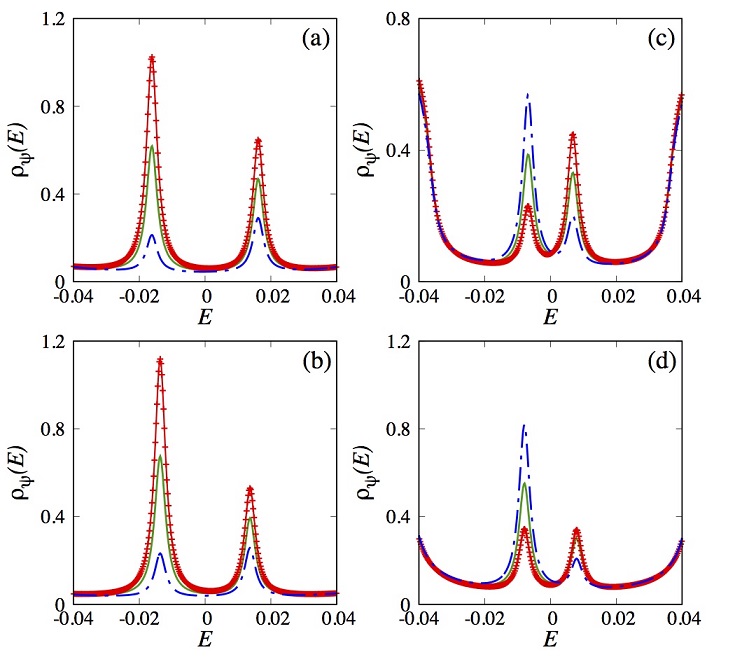}
	\caption{(color online) LDOS versus energy $E$ for the case with the third hybridization $\bar{V}_{\bm{k}}^{(3)}$. (a) $s$-wave pairing, (b) extended $s$-wave pairing, (c) $p$-wave pairing, (d) $d$+$id$ pairing. The green lines, red lines labeled with `$+$', and the blue dashed lines present the LDOS for $\bm{R}=(-1/\sqrt{3},0)$, $\bm{R}=(1/2\sqrt{3},1/2)$, and $\bm{R}=(1/2\sqrt{3},-1/2)$. The parameters for numerical calculation are given in Fig. \ref{fig7}.}%
	\label{fig8}%
\end{figure}

Fig. \ref{fig6} show the results for the second hybridization given in Eq. (\ref{eq22}).  For both these two kinds of hybridizations, one can find that the results for $p$-wave pairing and $d$+$id$ pairing exhibit only quantitative differences. Here we present these similarities.  At the position close to the impurity, $\bm{R}=(0,0)$, as shown in Figs. \ref{fig5}(c), \ref{fig5}(d), \ref{fig6}(c) and \ref{fig6}(d), there is only one resonance peak below the Fermi surface (see the red lines labeled with `$\times$' in these figures), at the nearest site close to the impurity, $\bm{R}=(-1/\sqrt{3},0)$, there are two resonance peaks located at the two sides of the Fermi energy (see the blue lines in the figures). The intensity of the other peak at positive energy in Fig. \ref{fig6}(c) is very small, (see the insert for more details). For topological nontrivial pairings ($p$-wave and $d$+$id$), the two kinds of hybridizations shown in Figs. \ref{fig5} and \ref{fig6} may be distinguishable, i.e., the maximum intensity of the negative resonance energy is located at the impurity site for the first hybridization, it is located on the bonds link the nearest neighbors of the impurity for the second hybridization.

Now we analysis the results for the third hybridization given in Eq. (\ref{eq23}). The first column in Fig. \ref{fig7} shows the LDOSs versus energy for different pairing symmetries and at different positions. One can find that, in this case, all of the four kinds of pairing symmetries exhibit two in-gap resonance peaks at the original point, $\bm{R}=(0,0)$ (see the red lines labeled with `$\times$' in Figs. \ref{fig7}(a)-\ref{fig7}(d) and inserts therein).
Another important difference is that, as shown in Figs. \ref{fig7}(e)-\ref{fig7}(k), the spacial distributions of the LDOSs break both the threefold rotation symmetry along the $z$-axis and the twofold rotation symmetry along the $y$-axis. Like the first two cases, the spacial distributions for the $p$-wave pairing and the $d$+$id$ pairing are very similar to each other, which demonstrates that the $p$-wave pairing and the $d$+$id$ pairing are difficult to be distinguished via the LDOS measurement of magnetic impurity resonance states. However, the spacial distributions of $p$-wave pairing and the $d$+$id$ pairing at the negative resonance energy is significant different from the $s$-wave pairing and the extended $s$-wave pairing, i.e., the maximum intensity points for the latter are located at $\bm{R}=(0,0)$, for the former, they are located at $\bm{R}\approx(0.25,-1.0)$.

Fig. \ref{fig8} give another formulation of the threefold rotation symmetry breaking. We calculate the LDOS versus energy at three rotation symmetric points for the four kinds of pairing potentials. One can find that the intensities of the resonance peaks are significant different for the three different locations.

\begin{figure*}[t]
	\includegraphics[width=\textwidth]{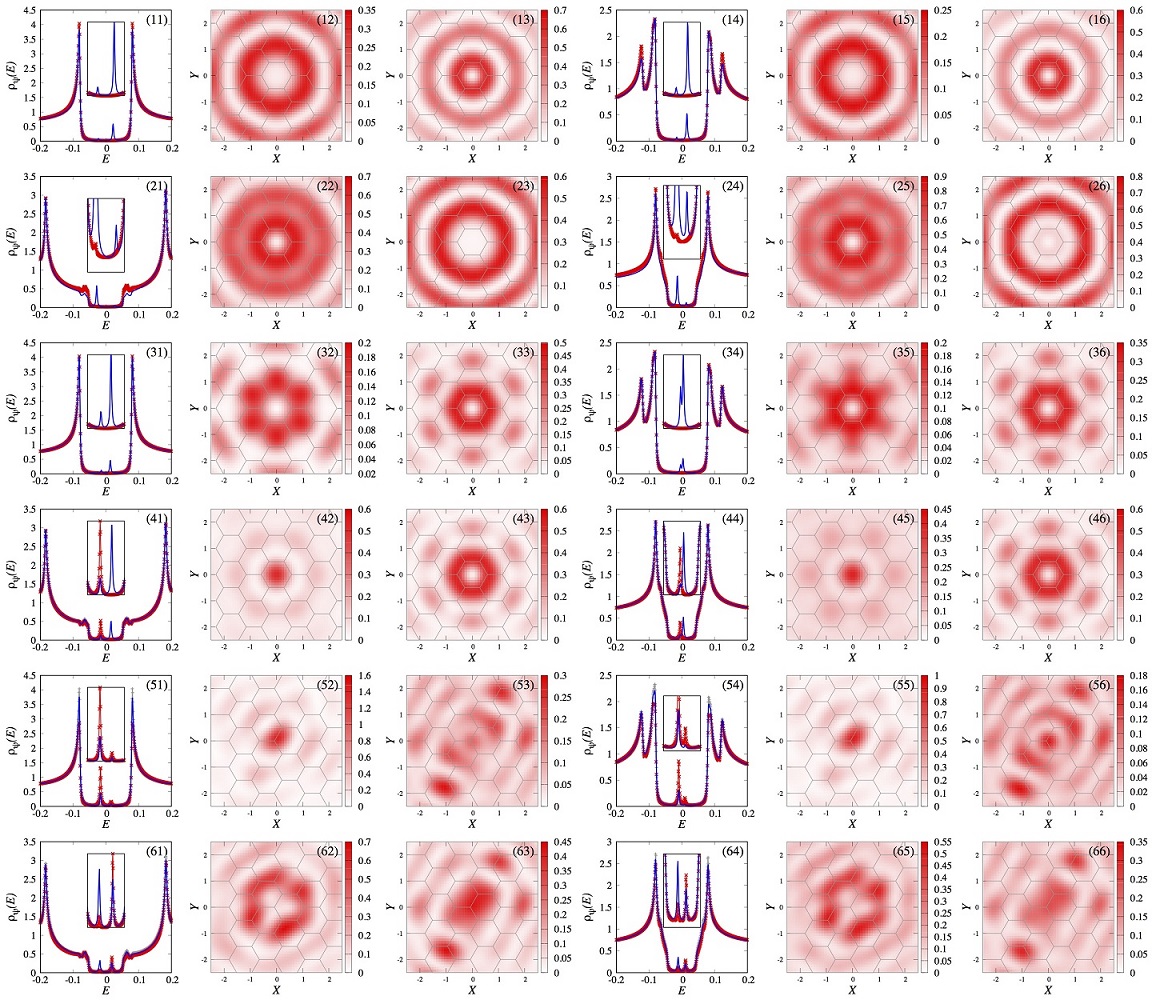}
	\caption{(color online) Integrated presentation of the LDOSs for $\mu=1.3058$ (the Van Hove singularity close to filling ratio $=1/2$). The first two rows show the results for the first hybridization $\bar{V}_{\bm{k}}^{(1)}$. The two rows in the middle show the results for the second hybridization $\bar{V}_{\bm{k}}^{(2)}$. The last two rows show the results for the third hybridization $\bar{V}_{\bm{k}}^{(3)}$. (11), (31) and (51) show LDOSs versus energy for the $s$-wave pairing with different hybridizations. (14), (34) and (54) show LDOSs versus energy for the extended $s$-wave pairing with different hybridizations. (21), (41) and (61) show LDOSs for the $p$-wave pairing. (24), (44) and (64) show the results for $d$+$id$ pairing. In each line plot, three situations are presented. The gray lines labeled with `$+$' show the clean LDOSs without impurity. The red lines labeled with `$\times$' show LDOSs at $\bm{R}=(0,0)$. The blue lines show LDOSs at the nearest neighbor, $\bm{R}=(-1/\sqrt{3},0)$. The contour plots close to the line plot show the topography of the two resonance states, e.g., (12) and (13) show the spacial distributions of the two in-gap resonance peaks in (11) at $E=-0.023$ and $E=0.023$, respectively. The inserts show the details of the resonance peaks.}%
	\label{fig9}%
\end{figure*}

We have also calculated LDOSs for different chemical potentials, $\mu=-1.6945$ (the Van Hove singularity close to filling ratio $=-1/2$), $\mu=1.5$ (filling ratio $=1/2$), and $\mu=1.3058$ (the other Van Hove singularity). We find that the results (i.e., the LDOSs versus energy and spacial distributions of the two in-gap resonance states) for $\mu=-1.6945$ are very close to that for $\mu=-1.7$, and the results for  $\mu=1.3058$ are qualitatively identical to that for $\mu=1.5$. Which demonstrate that the Van Hove singularities do not have significant influence on LDOS of resonance states induced by the magnetic impurity. Fig. \ref{fig9} gives an integrated presentation of LDOSs for $\mu=1.3058$. We find that the central results obtained for $\mu=-1.7$ still hold: 1) when the impurity is hybridized to the two orbitals at one site of the emergent honeycomb lattice, the spacial distributions shown in the last two rows in Fig. \ref{fig9} break both the threefold rotation symmetry along the $z$-axis and the twofold rotation symmetry along the $y$-axis, when the impurity is located at the center of the emergent honeycomb lattice and hybridized to the nearest neighbors symmetrically, the spacial distributions of the resonance states exhibit sixfold rotation symmetry, 3) for the impurity location given in 2), the LDOS versus energy for $s$-wave pairing and extended $s$-wave pairing at $\bm{R}=(0,0)$ do not have in-gap resonance peak, and 4) for the impurity location given in 2), the LDOS versus energy for $p$-wave pairing and $d$+$id$ pairing at $\bm{R}=(0,0)$ have only one in-gap resonance peak below the Fermi energy. The maximum point in the spacial distributions of the resonance states need to be emphasised here. For the third hybridization, $\bar{V}_{\bm{k}}^{(3)}$ given in Eq. (\ref{eq23}), the spacial distributions of the LDOSs show similar characteristics for different chemical potentials, see, i.e., Fig. \ref{fig7} and the last two rows of Fig. \ref{fig9} for details. For the other two hybridizations, however, the results are different. Comparing all the contour plots in Figs. \ref{fig5} and \ref{fig6}, one can find that there are only two subfigures where the maximum points in the spacial distributions of the resonance states are located at $\bm{R}=(0,0)$, Figs. \ref{fig5}(g) and \ref{fig5}(h), the resonance states corresponding to the negative resonance energies with $\bar{V}_{\bm{k}}^{(1)}$ hybridization for $p$-wave and $d$+$id$ pairing symmetries. When the chemical potential is doped to close to the other half filling ratio $1/2$, i.e., $\mu=1.308$ shown in Fig. \ref{fig9}, we find that in which the maximum points located at $\bm{R}=(0,0)$ are Figs. \ref{fig9}(42) and \ref{fig9}(45). Both of them correspond to the $\bar{V}_{\bm{k}}^{(2)}$ hybridization. Experimentally, when the impurity is located at the center of the emergent honeycomb lattice, the maximum point located at $\bm{R}=(0,0)$ is a strong evidence for unconventional pairing symmetry, however, which one of the two hybridizations is preferred for a specified magnetic impurity is difficult to discriminate. Our calculations suggest that, if the pairing symmetry is $p$-wave or $d$+$id$, by tuning the chemical potential close to the two half filling ratios $\pm1/2$, there must be a situation where the maximum point in the spacial distributions of the resonance states is located at $\bm{R}=(0,0)$ for the negative resonance energy.

Now we consider the influence of the term proportional to $t_5^{\prime\prime}$ given in Eq. (\ref{eq12}). We calculate the LDOS versus energy at $\bm{R}=(0,0)$ for (1) three different values of $t_5^{\prime\prime}$, $t_5^{\prime\prime}=0$, $t_2$, and $2t_2$, (2) two different topological nontrivial pairing, the $p$-wave pairing and $d$+$id$ pairing, (3) the three kind of hybridizations, and (4) the four different chemical potentials: filling ratio $=\pm1/2$ and the Van Hove singularities nearby (the plots are highly similar to each other, not shown here). For all the cases, we find that the LDOSs of the resonance states are almost not changed by the variation of  $t_5^{\prime\prime}$, so we suspect that the effect of the $t_5^{\prime\prime}$ term is limited.

\section{Conclusion}\label{sec4}

Based on the two orbital model proposed in Ref. \cite{YuanNFQ2018PRB}, we give a systematic study of the resonance states near a magnetic impurity in superconducting TwBG for typical pairing symmetries. We need to emphasize that though the parameters chosen in this work is different from those given in Ref. \cite{YuanNFQ2018PRB} (especially for the case $t_5^{\prime\prime}\ne0$ and $t_1^{\prime}=t_2^{\prime}=0$ used to fit the band structure given in Ref. \cite{NamNNT2017PRB}), our results are general and available for the parameters given in Ref. \cite{YuanNFQ2018PRB}.  Here are our main results from both symmetry analysis and numerical calculations:

1). When the impurity is hybridized to the two orbitals at one site of the emergent honeycomb lattice, i.e., the impurity is located at the center of the AB or BA spot, for any pairing symmetry belongs to the irreducible representations of $D_3$ point group, the spacial distribution of the resonance states will break both the threefold rotation symmetry around the $z$-axis and the twofold rotation symmetry around the $y$-axis.

2). When the impurity is located at the center of the emergent honeycomb lattice, i.e., at the center of the AA spot, and hybridized to the six nearest neighbors symmetrically, there are only two kinds of hybridizations which belong to the two-dimensional irreducible representations of the $D_3$ point group. For each hybridization and the four pairing symmetries studied, the spacial distributions of the resonance states reveal sixfold rotation symmetry.

3). For the hybridizations given in point 2), the in-gap resonance peak at the position close to the impurity must be vanishing for the $s$-wave pairing, extended $s$-wave pairing, and any other pairing symmetry belongs to the $\mathcal{A}_1$ representation of $D_3$ group.

4). For the hybridizations given in point 2), the unique resonance peak with negative resonance resonance energy at the position close to the impurity indicate that the pairing potential is topological nontrivial $p$-wave pairing or $d$+$id$ pairing.

Here we give a discussion about these results, focusing on further investigations. Firstly, these conclusions, especially point 1) and point 2), are essentially dependent on the assumption given in Ref. \cite{YuanNFQ2018PRB} that the two orbitals belong to the two-dimensional irreducible representation of $D_3$ group. Further investigations based on other models \cite{Bistritzer2011PNAS,Santos2007PRL,Santos2012PRB,MoonP2013PRB,TarnopolskyG2018arXiv1808,VenderbosJWF2018arXiv1808} will be important reference to find the pairing symmetry of superconducting magic angle TwBG. Secondly, the pairing symmetries we considered here are not complete.  When the inter-orbital pairing is considered, there will be lots of possible pairing potentials. For the on-site pairing, it is easy to check that there are 2 kinds of pairing potentials for each irreducible representation. For the nearest neighbor pairing and next to nearest neighbor pairing, there are numerous pairing potentials for each irreducible representation, a complete analysis for each pairing potential will be straightforward. Thirdly, as shown in the inserts of Figs. \ref{fig6}(c), \ref{fig6}(d), \ref{fig9}(21) and \ref{fig9}(24), even if the pairing potential is $p$-wave or $d$+$id$ symmetric, the in-gap resonance peak at $\bm{R}=(0,0)$ may be too weak to be detected. Fortunately, the in-gap resonance peak is significant for the charge density close to the other half-filling, see the inserts of Figs. \ref{fig9}(41), \ref{fig9}(44), \ref{fig5}(c) and \ref{fig5}(d), correspondingly. This is experimental accessible by tuning the gate voltage. Fourthly, when the impurity is located at the center of the emergent honeycomb lattice, the two hybridizations are obtained from symmetry analysis, for a practical adatom like manganese or chromium, the hybridization may break the rotation symmetries beforehand. Systematic theoretical investigations based on first-principle calculations may help to find the proper adatoms with desired hybridizations.

\section*{Acknowledgment}\label{sec5}
We appreciate the support from the NSFC under Grants No. 11504106 and No. 11447167 and the Fundamental Research Funds for the Central Universities under Grant No. 2018MS049.






\bibliographystyle{apsrev4-1}
\bibliography{magImp}

\end{document}